\documentclass[useAMS,usenatbib,usegraphicx]{mn2e}

\usepackage{natbib,amssymb,amsmath,amsfonts}
\bibpunct{(}{)}{;}{a}{}{,}

\usepackage{color,graphicx}
\graphicspath{{./}{./figs/}}

\usepackage{txfonts}
 \newcommand{\RM}[1]{\mathrm{#1}}         % math roman
 \newcommand{\V}[1]{\boldsymbol{#1}}      % vector
 \newcommand{\M}[1]{\mathbf{#1}}          % matrix
 \newcommand{\I}{\mathrm{i}}              % imaginary
 \newcommand{\E}{\mathrm{e}}              % exponential
 \renewcommand{\Re}{\operatorname{Re}}    % real part
 \renewcommand{\Im}{\operatorname{Im}}    % imaginary part
 \newcommand{\Var}{\operatorname{Var}}    % variance
\newcommand{\abs}[1]{\vert #1\vert}
\newcommand{\Abs}[1]{\left\vert #1\right\vert}

\newcommand{\bydef}{\stackrel{\mathrm{def}}{=}}

\newcommand{\Eq}[1]{Eq.~(\ref{#1})}
\newcommand{\Fig}[1]{Fig.~\ref{#1}}

\newcommand{\PSF}{\operatorname{PSF}}
\newcommand{\OTF}{\operatorname{OTF}}
\newcommand{\dyn}{\operatorname{dyn}}
\newcommand{\Nph}{N_\mathrm{ph}}         % number of photons

\usepackage{xspace}
\newcommand{\bkp}{\!\!\!}
\newcommand{\bkl}{\!\!\!\!\!}

\begin{document}

\title[High dynamic range imaging with a single-mode pupil remapping
  system]{High dynamic range imaging with a single mode pupil
  remapping system : a self calibration algorithm for redundant
  interferometric arrays}

\author[S. Lacour, E. Thi\'ebaut and G. Perrin]%
{S.~Lacour,$^1$ E.~Thi\'ebaut$^2$ and G.~Perrin$^1$ \\
$^1$ Observatoire de Paris -- Laboratoire d'Etudes Spatiales et
d'Instrumentation en Astrophysique, UMR-8109, 5 place Jules Janssen,
F-92195 Meudon, France \\
$^2$ Centre de Recherches Astronomiques de Lyon, UMR-5574, 9 avenue
Charles Andr\'e; F-69561 Saint Genis Laval Cedex}

\date{Accepted 2006 October 16. Received 2006 October 13; in original form
2006 September 13}

\maketitle

%%%%%%%%%%%%%%%%%%%%%%%%%%%%%%%%%%%%%%%%%%%%%%%%%%%%%%%%%%%%%
\begin{abstract}
The correction of the influence of phase corrugation in the pupil
plane is a fundamental issue in achieving high dynamic range
imaging. In this paper, we investigate an instrumental setup which
consists in applying interferometric techniques on a single telescope,
by filtering and dividing the pupil with an array of single-mode
fibers. We developed a new algorithm, which makes use of the fact
that we have a redundant interferometric array, to completely
disentangle the astronomical object from the atmospheric perturbations
(phase and scintillation). This self-calibrating algorithm can also be
applied to any -- diluted or not -- redundant interferometric setup.
On an 8 meter telescope observing at a wavelength of 630 nm, our
simulations show that a single mode pupil remapping system could
achieve, at a few resolution elements from the central star, a raw dynamic
range up to $10^6$; depending on the brightness of the
source.  The self calibration algorithm proved to be very efficient,
allowing image reconstruction of faint sources (mag = 15) even though
the signal-to-noise ratio of individual spatial frequencies are of the
order of 0.1.  We finally note that the instrument could be more
sensitive by combining this setup with an adaptive optics system. The
dynamic range would however be limited by the noise of the small, high
frequency, displacements of the deformable mirror.
\end{abstract}


%%%%%%%%%%%%%%%%%%%%%%%%%%%%%%%%%%%%%%%%%%%%%%%%%%%%%%%%%%%%%
\section{Introduction}
\label{sect:intro}  % \label{} allows reference to this section

The image obtained though a telescope is a convolution between the
brightness distribution of the astrophysical object and the point
spread function (PSF). In the Fourier domain, it is the multiplication
of the Fourier transform of the object and the Optical Transfer
Function (OTF). To restore a correct image of the source, one
therefore needs to know precisely the OTF. In the presence of static
aberrations only, deconvolution is possible since the OTF can be
obtained by observing an unresolved object. But when the OTF is
changing with time -- for example, in the presence of atmospheric
turbulence --, calibration requires averaging the perturbations, whose
parameters vary with time. This is one of the reasons why speckle
interferometry \citep{1970A&A.....6...85L}, one of the most well known
post-processing techniques, still has some difficulty to create high
dynamic range maps.

This mainly explains why real-time adaptive optics (AO) systems are a
fundamental feature of large telescopes. With such systems, the OTF of
the telescope is controlled by the deformable mirror to be the same as
the one of an uncorrupted telescope. However, technological limits
appear for i) larger telescopes (e.g.  extremely large telescopes),
ii) shorter wavelengths (e.g. visible), or iii) extremely high dynamic
range imaging (extreme adaptive optics). In these three cases it may
be advantageous to contemplate a complementary approach using
post-detection techniques. The combination of both could be the
solution to reach major scientific results like extra-solar planetary
system imaging. However, to do so, such techniques would require the
knowledge of the time varying OTF.

In \citet{2006..Perrin}, we proposed a passive solution (i.e.,
requiring no real-time modification of the optical path) by using a
remapping of the pupil. Single-mode fibers provide us with the
technology allowing such a massive modification of the geometry of the
pupil, while keeping zero optical path differences. In addition, they
also provide perfect spatial filtering. Data collection and analysis
are then similar to those utilized for aperture masking
\citep{1987Natur.328..694H,2000PASP..112..555T}, with the noticeable
advantage of having the flux of the whole entrance pupil, and the
possibility to completely disentangle instrumental from astrophysical
information.

In Sect.~\ref{sec:illpos} we explain why imaging through turbulence is
an ill-posed problem. After a recall of the principle of the
instrument, we show in Sect.~\ref{sec:principle} how what was before
an ill-posed problem can become a well-posed one.  This translate into
an algorithm described in Sect.\ref{sec:selfcal}. Finally, we show in
the simulations of Sect~\ref{sec:dyna} that we can therefore
reconstruct perfect images with a dynamic range only limited by
detector and photon noise. In Sect.~\ref{sec:conc} we conclude by
giving a brief summary of our results.

\section{The ill-posed problem of imaging through turbulence}
\label{sec:illpos}

\begin{figure*}
  \centering
  \resizebox{\hsize}{!}{\includegraphics{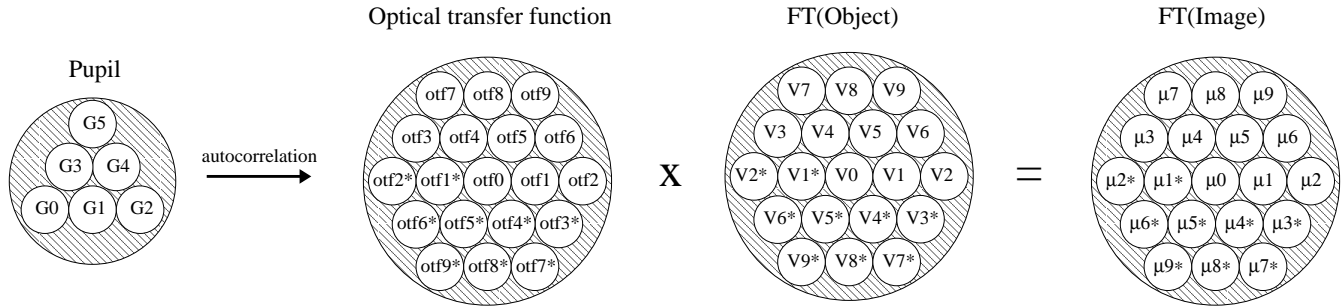}}
  \caption{ \label{fig:principe_sans} This sketch illustrates
    Eqs.~(\ref{eq:otf}) and (\ref{eq:muk}). The OTF result from the
    autocorrelation of the pupil complex amplitude transmission,
    and the Fourier transform of the image is the multiplication of
    the OTF by the Fourier transform of the object observed. The
    unknowns are the 15 complex values
    $\{G_O,\ldots,G_5,V_1,\ldots,V_9\}$, whereas the observables
    provide only 9 complex values
    $\{\mu_1,\ldots,\mu_9\}$. Deconvolution is therefore an ill-posed
    problem.}
\end{figure*}

The image formed in the focal plane of a telescope is the convolution
of the object brightness distribution $O(\V{x})$ with the point
spread function $\PSF(\V{x})$ of the instrument:
\begin{equation}
  I(\V{x}) = O(\V{x}) * \PSF(\V{x}).
\end{equation}
In the Fourier domain, the convolution operation is transformed into a
multiplication, while the Fourier transform of the point spread
function is the optical transfer function (OTF):
\begin{equation}
  \mu(\V{u}) = V(\V{u}) \times \OTF(\V{u}).
  \label{eq:base}
\end{equation}
We choose $\mu(\V{u})$ as the Fourier transform of the image, and
$V(\V{u})$ as the Fourier transform of the object brightness
distribution. This is to be in line with interferometric conventions,
where it is also called the visibility function. The fact that the
image depends on two unknown functions, $V(\V{u})$ and $\OTF(\V{u})$,
is the problem underlying any image reconstruction algorithm; without
adding further information, we have no way to disentangle the object
from the PSF.

Following an interferometric approach, we discretize the OTF to reduce
the problem to a system of observables and unknowns. The OTF results
from the autocorrelation of the complex values of the complex
amplitude transmission inside the pupil. Thus, the OTF can be
discretized by considering the pupil as being made of a number of
coherent patches where phase and amplitude variations are negligible.
Each patch is defined by a position vector $\V{r}_{i}$ and a
complex amplitude transmission:
\begin{equation}
  G(\V{r}_{i}) = g_i \, \E^{\I\,\phi_i}
  \label{eq:Gi}
\end{equation}
with a phase $\phi_i$ (e.g. atmospheric piston), an amplitude $g_i$
(e.g. scintillation) and where $\I^2{\,\bydef}-1$. Each
pair of patches $(i,j)$ selects one specific spatial frequency
described by the frequency vector
$\V{u}_{k}=(\V{r}_{i}-\V{r}_{j})/\lambda$; where $\V{r}_{i}$ and
$\V{r}_{j}$ are the location vectors of the patches projected in a
plane perpendicular to the line of sight and $\lambda$ is the
wavelength. Hence, the OTF at frequency vector $\V{u}_{k}$ is obtained
by the relation:
\begin{equation}
  \OTF(\V{u}_k)=\sum_{(i,j)\in\mathcal{B}_k}
  G(\V{r}_{i}) \, G(\V{r}_{j})^\star,
  \label{eq:otf}
\end{equation}
where $\mathcal{B}_k$ is the set of
aperture pairs which sample the $k$-th
spatial frequency $\V{u}_k$:
%$\V{u}_{k}= (\V{r}_{i}-\V{r}_{j})/\lambda$:
\begin{equation}
  \mathcal{B}_k =
  \Bigl\{\,(i,j) \ :\ 
  %\mathrm{\quad s.t.\quad}
  (\V{r}_i - \V{r}_j)/\lambda = \V{u}_k
  \,\Bigr\}
  \label{eq:set-of-pairs}
\end{equation}
This shows that the optical transfer function can be obtained from the
knowledge of the complex amplitude transmission inside the pupil.
Using \Eq{eq:base}, we can deduce a direct relation between the pupil
transmission, the Fourier transform of the image, and the Fourier
transform of the brightness distribution of the object:
\begin{equation}
  \mu(\V{u}_{k}) = V(\V{u}_{k})
  \!\!\sum_{(i,j)\in\mathcal{B}_k}\!\!
  G(\V{r}_{i})\,
  G(\V{r}_{j})^\star.
  \label{eq:mukl}
\end{equation}
Or to simplify the notation:
\begin{equation}
  \mu_{k} = {V}_{k} \sum_{(i,j)\in\mathcal{B}_k} \!\! G_i\, G_j^\star
  \,,
  \label{eq:muk}
\end{equation}
where, and hereinafter, we define:
$\mu_k{\,\bydef\,}\mu(\V{u}_{k})$,
$V_k{\,\bydef\,}V(\V{u}_{k})$,
$\OTF_k{\,\bydef\,}\OTF(\V{u}_{k})$ and
$G_i{\,\bydef\,}G(\V{r}_{i})$.

The image reconstruction problem is then reduced to finding the
unknowns $\{V_k,G_i;\forall k,\forall i\}$ given the $\mu_k$'s.  The
ill-posedness of this task can be exhibited thanks to a simple
example.  In \Fig{fig:principe_sans}, the complex amplitude
transmission in the pupil is binned into six different elements (the
$G_i$'s). The autocorrelation of these six patches creates an OTF
defined by a real value $\OTF_0$ (at central spatial frequency) and 9
complex values $(\OTF_1,\ldots,\OTF_9)$.  These OTF values multiplied
by the visibility function of the astronomical object
$(V_0{\equiv}1,V_1,\ldots,V_9)$ yield the Fourier transform of the
observed image as one real value $\mu_0$ and 9 complex values
$(\mu_1,\ldots,\mu_9)$.  Since, by definition, the real value $V_0$ is
equal to 1 and since the phase of one of the complex amplitude
transmissions can be arbitrarily chosen, the image reconstruction
involves the computation of 29 unknowns (15 complex values:
$G_O,\ldots,G_5,V_1,\ldots,V_9$, minus an arbitrary phase) given only
19 measurements (the real value $\mu_0$ and the 9 complex values
$\mu_1,\ldots,\mu_9$).  Our example demonstrates that the image
reconstruction when the PSF is unknown is an ill-posed problem termed
as \emph{blind deconvolution} \citep{Thiebaut_Conan-1995-bdec}.
Without adding further information, disentangling astronomical from
instrumental information is impossible.

To avoid having to disentangle the time-dependent OTF, a traditional
solution is to average its fluctuations.  Over multiple observations,
the long exposure OTF is:
\begin{equation}
  \OTF_k=\left\langle
  \sum_{(i,j)\in\mathcal{B}_k}\!\!
  G_i\,G_j^\star
  \right\rangle\,.
  \label{eq:lp}
\end{equation}
To calibrate the OTF, the astronomer can observe a point-like star
(i.e. such that $V_k=1, \forall k$) and apply the same averaging
process. However, when phase variations become larger than wavelength,
the average of the complex OTF tends toward 0, and deconvolution
is impossible with a finite S/N ratio \citep{Thiebaut-2005-Cargese}.
In practice, long exposure images have a $\lambda/r_0$ effective
resolution, where $r_0\simeq20\,\mathrm{cm}$ in the visible is Fried's
parameter.  Two solutions have been proposed to overcome this problem
and achieve the diffraction limit at $\lambda/D$ where $D$ is the
pupil diameter.  The first solution is to correct the wavefront in
real time so as to keep the wavefront perturbations smaller than the
wavelength.  This is achieved with an \emph{adaptive optics} system.
The second solution, so called \emph{speckle interferometry}
\citep{1970A&A.....6...85L}, is to take short exposures with respect to the
time scale of the perturbation, and to average the squared modulus of
the Fourier transform of the image.  This way, the transfer function
for the modulus of the Fourier transform of the observed brightness
distribution becomes:
\begin{equation}
  \OTF_k = \sqrt{\Bigl\langle\,
    \Bigl\vert\sum_{(i,j)\in\mathcal{B}_k}\!\!
    G_i \, G_j^\star\Bigr\vert^2\,
    \Bigr\rangle}
  \label{eq:sp}
\end{equation}
and is attenuated for spatial frequencies higher than $r_0/\lambda$
but different from zero up to $D/\lambda$.  The Fourier phase of the
observed brightness distribution can be retrieved by means of a third
order technique such as the bispectrum \citep{Weigelt1977}.

Here we propose an alternative approach. Instead of averaging the OTF,
the goal is to have real time measurements of the complex
amplitude transmission. Then, a post-detection algorithm can be used
to obtain a calibrated OTF:
\begin{equation}
  \OTF_k = \left\langle
  \sum_{(i,j)\in\mathcal{B}_k}
  \frac{G_i \, G_j^\star}{\widetilde{G}_i \, \widetilde{G}_j^\star}
  \right\rangle
  \label{eq:rp}
\end{equation}
where $\widetilde{G}_i$ and $\widetilde{G}_j$ are estimated
complex amplitude transmissions.  We have however demonstrated
in this section that this information is unavailable on a simple image
of the object.  In order to recover the missing information, we have
proposed a system \citep{2006..Perrin}, in which the telescope pupil
is injected into an array of single-mode fibers, and rearranged into a
new non-redundant exit pupil.

\section{From an ill-posed to a well-posed problem}
\label{sec:principle}

\subsection{The instrument}
\label{sec:instru}

\begin{figure}
  \centering
  \includegraphics[width=7.5cm]{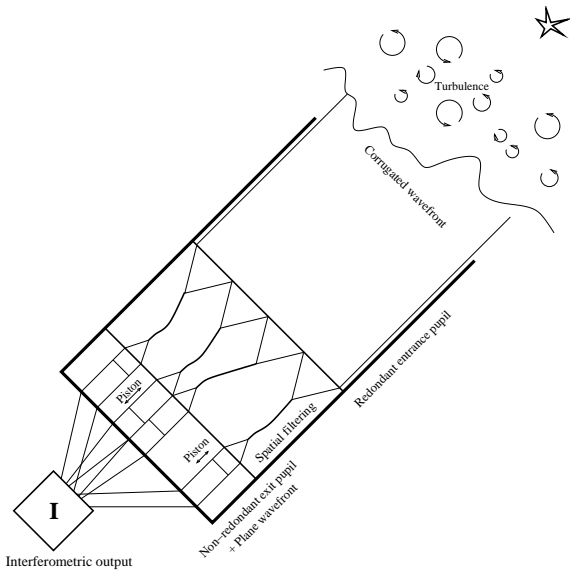}
  \caption{ \label{fig:concept} In this instrument, the pupil (or an
    image of it) is subdivided into several sub-pupils whose outputs are
    injected into single-mode fibers. The fibers are then rearranged
    to create a new non-redundant pupil. Imaging on the detector is
    then performed as if no remapping had taken place.}
\end{figure}

The concept, proposed in \citet{2006..Perrin}, is summarised in
\Fig{fig:concept}. Briefly, entrance sub-apertures collect
independently the radiation from an astronomical source in the pupil
of the telescope, and focus the light onto the input heads of
single-mode optical fibers (of location vectors $\V{r}_{i}$). The
radiation is then guided by the fibers down to a recombination unit,
in which the beams are rearranged into a 1D or 2D non-redundant
configuration to form the exit pupil. Finally, the remapped output
pupil is focused to form fringes in the focal plane where a different
fringe pattern is obtained for every pair of sub-pupils.

The amplitudes and phases of the fringes are measurements of the
Fourier components given by the entrance baselines vectors
$(\V{r}_{i}-\V{r}_{j})$; the Fourier components measured in
the image are thus given by the relation:
\begin{equation}
  \mu_{i,j} = V_k \, G_i \, G_j^\star \, ,
  \label{eq:mul_l}
\end{equation}
where $V_k$ is the complex visibility of the observed object at the
frequency $\V{u}_{k}=(\V{r}_{i}-\V{r}_{j})/\lambda$,
and $G_i$ and $G_j$ are the complex transmission factors in the
telescope pupil as defined in \Eq{eq:Gi}.  It is interesting to
note the differences between this relation and \Eq{eq:muk}: thanks to
the remapping, each measurement now corresponds to a single pair of
sub-apertures.

A second advantage of this setup comes from the fact that single-mode
fibers act as spatial filters. As a consequence, the relation
$G_i=g_i\,\E^{\I\,\phi_i}$ is exact for each sub-pupil. Indeed,
after being filtered by the fiber, the complex electric field, which
is otherwise a continuous function, can be characterized by only two
parameters: its phase and its amplitude.  The discretization
introduced in the previous section is no longer an approximation which
opens the way toward searching for an exact solution.

\subsection{On the unicity of the solution}
\label{sec:why_remap}

\begin{figure*}
  \centering
  \resizebox{\hsize}{!}{\includegraphics{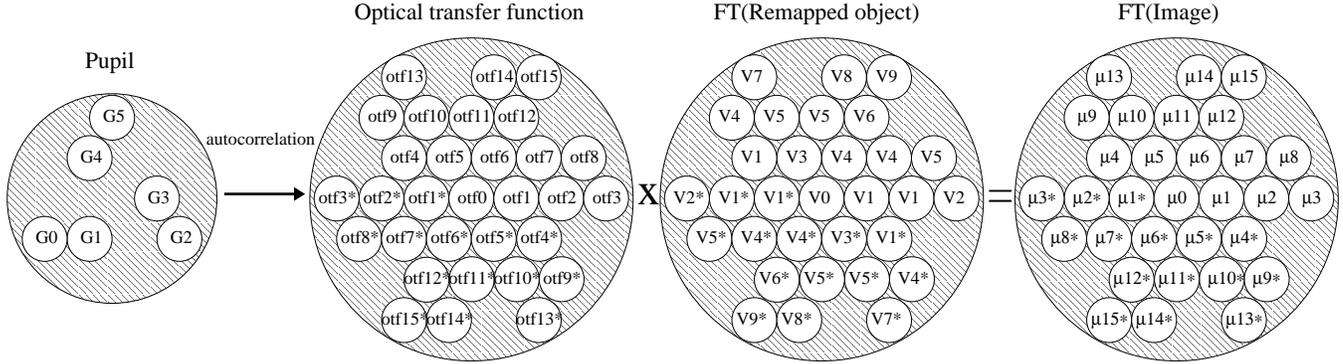}}
  \caption{ \label{fig:principe_avec} This sketch illustrates
    Eqs.~(\ref{eq:otf}) and (\ref{eq:mul_l}) in the case of a remapped
    pupil. As in \Fig{fig:principe_sans}, the OTF result from the
    autocorrelation of the pupil complex amplitude transmission,
    and the Fourier transform of the image is the multiplication of
    the OTF by the visibility values of the observed object. However,
    whereas there are still 15 unknown complex values
    $\{G_O,\ldots,G_5,V_1,\ldots,V_9\}$, the observables provide 15
    complex values $\{\mu_,\ldots,\mu_{15}\}$ and deconvolution is
    therefore possible.}
\end{figure*}

The fundamental idea of this paper comes from the fact that by using
interferometric techniques, information can be retrieved to deconvolve
an image from its PSF. As stated in Sect.~\ref{sec:illpos}, image
restoration requires the knowledge of the complex transmission terms
$G_i$, which is impossible with direct imaging. However,
\citet{1982OptCo..42..157G} proved that the missing information
can be encoded at higher frequencies \citep[see
also][]{1983OptCo..45..380A,1985Arnot.TI}.  Remapping enables an
increase in the number of observables $\mu$ while keeping the
number of unknowns constant.  This is possible since the complex
visibilities $V_k$ only depend on the baselines in the telescope
entrance pupil.  They do not change with a rearrangement of the pupil
\citep{1992A&A...253..641T}.

This can be well understood in terms of unknowns and observables.  A
remapped system is governed by \Eq{eq:mul_l}. Providing $M$
sub-apertures and $R$ redundant entrance baselines, the number of
complex unknowns are of $M (M-1)/2-R$ visibilities ($V_k$ terms), and
$M-1$ transmission factors ($G_i$ terms). On the other hand, the
number of measurements is $M(M-1)/2$ (the $\mu_{i,j}$ terms; with $i
\neq j$). Hence, if $R>M-1$, there are more observables than unknowns
and the system of equations can be solved.

The fact that both the $V_k$ and the $G_i$ terms can be deduced from
the $\mu_{i,j}$ can be illustrated in a specific
case. \Fig{fig:principe_avec} is the same sketch as in
\Fig{fig:principe_sans}, but with the 6 sub-pupils rearranged
into a non-redundant configuration. This configuration was chosen to
have the most compact configuration, but any other non-redundant
configuration could have been used \citep[see for
example][]{1970..Golay}.  On the left panel are the complex
transmission factors of the remapped pupil. The other panels show the
Fourier transform values of, from left to right, the PSF, the
astronomical object, and the image on the detector.  The equation
linking the observables $\mu_{i,j}$ to the unknowns $G_i$ and $V_{k}$
is \Eq{eq:mul_l}. Inversion of the resulting set of equations may
be possible since the number of unknowns is larger than the number of
measurements. This can be demonstrated by using the logarithm of terms
in \Eq{eq:mul_l}, which becomes
\begin{equation}
\ln(|\mu_{i,j}|) = \ln(|V_k|)+\ln(g_i)+\ln(g_j),
\end{equation}
for the real part, and
\begin{equation}
\Phi(\mu_{i,j}) = \Phi(V_k)+\phi_i-\phi_j,
\end{equation}
for the imaginary part. In these two equations, $\Phi()$ is the
argument function, and $g_i$, $g_j$, $\phi_i$ and $\phi_j$ are as
defined in \Eq{eq:Gi}. We obtain this way two sets of linear
equations, one for for the phases:
\begin{equation}
  \left[ \Phi(\mu) \right] = \M{M}_\RM{P} \cdot
  \left(
  \begin{array}{c}
   \left[  \phi \right] \\
   \left[  \Phi(V) \right]
  \end{array}
  \right)
  \label{eq:mp_short}
\end{equation}
and one for the amplitudes:
\begin{equation}
  \left[ \ln(|\mu|) \right] = \M{M}_\RM{A} \cdot
  \left(
  \begin{array}{c}
    \left[ \ln(g) \right] \\
    \left[ \ln(|V|) \right] \end{array}
  \right)  
\end{equation}
where $\left[ \ \right]$ represents column vectors.  $\M{M}_\RM{P}$
and $\M{M}_\RM{A}$ are two matrices containing 1, 0, and -1 values.
Specifically to the example of \Fig{fig:principe_avec},
\Eq{eq:mp_short} becomes: 
{ 
%\footnotesize
\scriptsize
%\tiny
  \begin{displaymath}
  \left(\!
  \begin{array}{l}
    \Phi(\mu_1) \\
    \Phi(\mu_2) \\
    \Phi(\mu_3) \\
    \Phi(\mu_4) \\
    \Phi(\mu_5) \\
    \Phi(\mu_6) \\
    \Phi(\mu_7) \\
    \Phi(\mu_8) \\
    \Phi(\mu_9) \\
    \Phi(\mu_{10}) \\
    \Phi(\mu_{11}) \\
    \Phi(\mu_{12}) \\
    \Phi(\mu_{13}) \\
    \Phi(\mu_{14}) \\
    \Phi(\mu_{15})
  \end{array}
  \!\right) = \left(
  \begin{array}{rrrrrrrrrrrrrrr}
    1&\bkl-1&\bkp0&\bkp0&\bkp0&\bkp0&\bkp1&\bkp0&\bkp0&\bkp0&\bkp0&\bkp0&\bkp0&\bkp0&\bkp0 \\
    0&\bkp1&\bkl-1&\bkp0&\bkp0&\bkp0&\bkp1&\bkp0&\bkp0&\bkp0&\bkp0&\bkp0&\bkp0&\bkp0&\bkp0 \\
    1&\bkp0&\bkl-1&\bkp0&\bkp0&\bkp0&\bkp0&\bkp1&\bkp0&\bkp0&\bkp0&\bkp0&\bkp0&\bkp0&\bkp0 \\
    0&\bkp0&\bkp0&\bkp1&\bkl-1&\bkp0&\bkp1&\bkp0&\bkp0&\bkp0&\bkp0&\bkp0&\bkp0&\bkp0&\bkp0 \\
    0&\bkp0&\bkp1&\bkl-1&\bkp0&\bkp0&\bkp0&\bkp0&\bkp1&\bkp0&\bkp0&\bkp0&\bkp0&\bkp0&\bkp0 \\
    0&\bkp0&\bkp0&\bkp0&\bkp1&\bkl-1&\bkp0&\bkp0&\bkp0&\bkp1&\bkp0&\bkp0&\bkp0&\bkp0&\bkp0 \\
    0&\bkp1&\bkp0&\bkl-1&\bkp0&\bkp0&\bkp0&\bkp0&\bkp0&\bkp1&\bkp0&\bkp0&\bkp0&\bkp0&\bkp0 \\
    1&\bkp0&\bkp0&\bkl-1&\bkp0&\bkp0&\bkp0&\bkp0&\bkp0&\bkp0&\bkp1&\bkp0&\bkp0&\bkp0&\bkp0 \\
    0&\bkp0&\bkp1&\bkp0&\bkl-1&\bkp0&\bkp0&\bkp0&\bkp0&\bkp1&\bkp0&\bkp0&\bkp0&\bkp0&\bkp0 \\
    0&\bkp0&\bkp0&\bkp1&\bkp0&\bkl-1&\bkp0&\bkp0&\bkp0&\bkp0&\bkp1&\bkp0&\bkp0&\bkp0&\bkp0 \\
    0&\bkp1&\bkp0&\bkp0&\bkl-1&\bkp0&\bkp0&\bkp0&\bkp0&\bkp0&\bkp1&\bkp0&\bkp0&\bkp0&\bkp0 \\
    1&\bkp0&\bkp0&\bkp0&\bkl-1&\bkp0&\bkp0&\bkp0&\bkp0&\bkp0&\bkp0&\bkp1&\bkp0&\bkp0&\bkp0 \\
    0&\bkp0&\bkp1&\bkp0&\bkp0&\bkl-1&\bkp0&\bkp0&\bkp0&\bkp0&\bkp0&\bkp0&\bkp1&\bkp0&\bkp0 \\
    0&\bkp1&\bkp0&\bkp0&\bkp0&\bkl-1&\bkp0&\bkp0&\bkp0&\bkp0&\bkp0&\bkp0&\bkp0&\bkp1&\bkp0 \\
    1&\bkp0&\bkp0&\bkp0&\bkp0&\bkl-1&\bkp0&\bkp0&\bkp0&\bkp0&\bkp0&\bkp0&\bkp0&\bkp0&\bkp1
  \end{array}
  \right) \cdot \left(\!
  \begin{array}{c}
    \phi_0 \\
    \phi_1 \\
    \phi_2 \\
    \phi_3 \\
    \phi_4 \\
    \phi_5 \\
    \Phi(V_1) \\
    \Phi(V_2) \\
    \Phi(V_3) \\
    \Phi(V_4) \\
    \Phi(V_5) \\
    \Phi(V_6) \\
    \Phi(V_7) \\
    \Phi(V_8) \\
    \Phi(V_9)
  \end{array}
  \!\right) \, .
  %\label{eq:matrix_P}
  \end{displaymath}
}
The rank of this matrix is 12, while the number of unknowns is
15. The three terms of degeneracy are one for the absolute phase
reference, and two for the tip and tilt. Thus, by providing an
arbitrary constrain on these three terms (the absolute phase is
arbitrary and the tip and tilt only depend on the location of the
image centroid), we can perform a singular value decomposition of the
matrix and obtain from the measurements $\mu_{i,j}$ a unique solution
for the phase of the perturbations and object visibilities. The same
method applies to the logarithm of the amplitude:
{
%\footnotesize
\scriptsize
%\tiny
  \begin{displaymath}
    \left(\!
    \begin{array}{l}
      \ln(|\mu_1|) \\
      \ln(|\mu_2|) \\
      \ln(|\mu_3|) \\
      \ln(|\mu_4|) \\
      \ln(|\mu_5|) \\
      \ln(|\mu_6|) \\
      \ln(|\mu_7|) \\
      \ln(|\mu_8|) \\
      \ln(|\mu_9|) \\
      \ln(|\mu_{10}|) \\
      \ln(|\mu_{11}|) \\
      \ln(|\mu_{12}|) \\
      \ln(|\mu_{13}|) \\
      \ln(|\mu_{14}|) \\
      \ln(|\mu_{15}|)
    \end{array}
    \!\right) = \left(
    \begin{array}{rrrrrrrrrrrrrrrr}
      1&\bkp1&\bkp0&\bkp0&\bkp0&\bkp0&\bkp1&\bkp0&\bkp0&\bkp0&\bkp0&\bkp0&\bkp0&\bkp0&\bkp0 \\
      0&\bkp1&\bkp1&\bkp0&\bkp0&\bkp0&\bkp1&\bkp0&\bkp0&\bkp0&\bkp0&\bkp0&\bkp0&\bkp0&\bkp0 \\
      1&\bkp0&\bkp1&\bkp0&\bkp0&\bkp0&\bkp0&\bkp1&\bkp0&\bkp0&\bkp0&\bkp0&\bkp0&\bkp0&\bkp0 \\
      0&\bkp0&\bkp0&\bkp1&\bkp1&\bkp0&\bkp1&\bkp0&\bkp0&\bkp0&\bkp0&\bkp0&\bkp0&\bkp0&\bkp0 \\
      0&\bkp0&\bkp1&\bkp1&\bkp0&\bkp0&\bkp0&\bkp0&\bkp1&\bkp0&\bkp0&\bkp0&\bkp0&\bkp0&\bkp0 \\
      0&\bkp0&\bkp0&\bkp0&\bkp1&\bkp1&\bkp0&\bkp0&\bkp0&\bkp1&\bkp0&\bkp0&\bkp0&\bkp0&\bkp0 \\
      0&\bkp1&\bkp0&\bkp1&\bkp0&\bkp0&\bkp0&\bkp0&\bkp0&\bkp1&\bkp0&\bkp0&\bkp0&\bkp0&\bkp0 \\
      1&\bkp0&\bkp0&\bkp1&\bkp0&\bkp0&\bkp0&\bkp0&\bkp0&\bkp0&\bkp1&\bkp0&\bkp0&\bkp0&\bkp0 \\
      0&\bkp0&\bkp1&\bkp0&\bkp1&\bkp0&\bkp0&\bkp0&\bkp0&\bkp1&\bkp0&\bkp0&\bkp0&\bkp0&\bkp0 \\
      0&\bkp0&\bkp0&\bkp1&\bkp0&\bkp1&\bkp0&\bkp0&\bkp0&\bkp0&\bkp1&\bkp0&\bkp0&\bkp0&\bkp0 \\
      0&\bkp1&\bkp0&\bkp0&\bkp1&\bkp0&\bkp0&\bkp0&\bkp0&\bkp0&\bkp1&\bkp0&\bkp0&\bkp0&\bkp0 \\
      1&\bkp0&\bkp0&\bkp0&\bkp1&\bkp0&\bkp0&\bkp0&\bkp0&\bkp0&\bkp0&\bkp1&\bkp0&\bkp0&\bkp0 \\
      0&\bkp0&\bkp1&\bkp0&\bkp0&\bkp1&\bkp0&\bkp0&\bkp0&\bkp0&\bkp0&\bkp0&\bkp1&\bkp0&\bkp0 \\
      0&\bkp1&\bkp0&\bkp0&\bkp0&\bkp1&\bkp0&\bkp0&\bkp0&\bkp0&\bkp0&\bkp0&\bkp0&\bkp1&\bkp0 \\
      1&\bkp0&\bkp0&\bkp0&\bkp0&\bkp1&\bkp0&\bkp0&\bkp0&\bkp0&\bkp0&\bkp0&\bkp0&\bkp0&\bkp1
    \end{array}
    \right) \cdot \left( \!
    \begin{array}{l}
      \ln(g_0) \\
      \ln(g_1) \\
      \ln(g_2) \\
      \ln(g_3) \\
      \ln(g_4) \\
      \ln(g_5) \\
      \ln(|V_1|) \\
      \ln(|V_2|) \\
      \ln(|V_3|) \\
      \ln(|V_4|) \\
      \ln(|V_5|) \\
      \ln(|V_6|) \\
      \ln(|V_7|) \\
      \ln(|V_8|) \\
      \ln(|V_9|)
    \end{array}
    \! \right) \, .
    \label{eq:matrix_A}
  \end{displaymath}}
The rank of this matrix is 14, meaning all the amplitudes can be
retrieved, except for the total brightness of the object. This
parameter can easily be constrained by normalizing the flux of the
reconstructed image. The measurement of the amplitudes is an
important issue since we have to correct for injection variability in
the single-mode fibers.

It is however clear that solving this system would require taking the
logarithm of the measurements. This would be very sensitive to
additive noise. To get the best out of the data, it is better to fit
the measurements using their complex values and \Eq{eq:mul_l}. To do
so, we developed a self-calibration algorithm which permits the use of
thousands of snapshot all together to reconstruct an image up to the
photon noise limit.

\subsection{A self-calibration algorithm for redundant arrays}
\label{sec:selfcal}

This section presents a self-calibration algorithm adapted to
single-mode pupil remapping instruments, but also more generally to
any kind of redundant interferometric array.  Indeed, the equation
$\mu_{i,j} = V_k \, G_i \, G_j^\star$ established in
Sec~\ref{sec:instru} is common to all interferometric facilities. In
the case of long baseline interferometry for example, $\mu_{i,j}$ is
the measurement of the complex coherence value between telescopes $i$
and $j$, $G_i$ the complex transmission factor of telescope $i$, and
$V_k$ the complex visibility of the astronomical object at the
baseline formed by telescope $i$ and $j$.

The particularity of this self-calibration algorithm comes from
the fact that it gives complex visibility estimations without the need
of a regularization term. This is possible thanks to the redundancy of
the interferometric array. If one wants to make sure this algorithm is
adapted to a specific interferometric facility, he would have first to
establish the $\M{M}_\RM{P}$ and $\M{M}_\RM{A}$ matrices, and thus
verify the unicity of the solution.

In the next sections, we first start deriving an algorithm in the
single-exposure case (Sect.~\ref{sec:selfcal1},~\ref{sec:selfcal2}
and~\ref{sec:self-calib}), and then we show how to extend our
algorithm to account for multiple exposures
(Sect.~\ref{sec:selfcal4},~\ref{sec:selfcal5}).

\subsubsection{Log-likelihood}
\label{sec:selfcal1}

Following the \citet{1985..Goodman} model for the noise of measured
complex visibilities, we assume that different measured complex
visibilities are uncorrelated and that, for a given measured complex
visibility $\mu_{i,j}$, the real and imaginary parts are uncorrelated
Gaussian random variables which have the same standard deviation.
Under these assumptions and from \Eq{eq:mul_l}, the log-likelihood of
the data is:
\begin{equation}
  \label{eq:chi2}
  \ell(\V{V},\V{G}) = \sum_k \sum_{(i,j)\in\mathcal{B}_k}\!\!
  w_{i,j}\,\left\vert\mu_{i,j} - G_i\,G_j^\star\,V_k\right\vert^2
\end{equation}
where $G_i$ and $G_j$ are the complex transmissions for each
sub-aperture and where $\mathcal{B}_k$ is the set of sub-aperture
pairs for which the interferences sample the $k$-th spatial frequency
$\V{u}_k$ as defined by \Eq{eq:set-of-pairs}.  In \Eq{eq:chi2}, the
statistical weights are:
\begin{equation}
  \label{eq:weight}
  w_{i,j}
  = \frac{1}{\Var\bigl(\Re(\mu_{i,j})\bigr)}
  = \frac{1}{\Var\bigl(\Im(\mu_{i,j})\bigr)}
  \,.
\end{equation}

Solving the image reconstruction problem, in the maximum likelihood
sense, consists in seeking for the complex transmissions $G_i$ and the
object visibilities $V_k$ which minimize the value of
$\ell(\V{V},\V{G})$ given by \Eq{eq:chi2}.  Unfortunately, the
log-likelihood $\ell(\V{V},\V{G})$ being a polynomial of 6th degree
with respect to the unknowns (the $V_k$'s and the $G_i$'s), proper
means to minimize it have to be invented.

\subsubsection{Best object visibilities}
\label{sec:selfcal2}

Given the complex transmissions $\V{G}$, the expression of
$\ell(\V{V},\V{G})$ in \Eq{eq:chi2} is quadratic with respect to the
object complex visibilities $\V{V}$.  Providing the complex
transmissions $\V{G}$ are known, obtaining the best object complex
visibilities $\V{V}$ is then a simple linear least-squares problem.
The solution of this problem is found by solving:
\begin{equation}
  \label{eq:1st-optim-cond-V}
  \frac{\partial\ell}{\partial V_k} = 0\,,\quad\forall k 
\end{equation}
where, by linearity, the derivative of the real quantity
$\ell(\V{V},\V{G})$ with respect to the complex $V_k$ is defined as:
\begin{equation}
  \label{eq:cmplx-deriv}
  \frac{\partial\ell}{\partial V_k} \bydef
  \frac{\partial\ell}{\partial \Re\left(V_k\right)}
  + \I\,\frac{\partial\ell}{\partial \Im\left(V_k\right)} \,.
\end{equation}
  Then:
\begin{eqnarray}
  \frac{\partial\ell}{\partial V_k}
  &=& 2\!\sum_{(i,j)\in\mathcal{B}_k}\!
      w_{i,j}\,\left(G_i\,G_j^\star\,V_k - \mu_{i,j}\right)\,G_i^\star\,G_j
      \nonumber\\
  &=& 2\,V_k\!\sum_{(i,j)\in\mathcal{B}_k}w_{i,j}\,\abs{G_i}^2\,\abs{G_j}^2
     -2\!\sum_{(i,j)\in\mathcal{B}_k}w_{i,j}\,\mu_{i,j}\,G_i^\star\,G_j\,.
     \label{eq:chi2-pder-Vk}
\end{eqnarray}
Solving \Eq{eq:1st-optim-cond-V} with the partial derivative expression in
\Eq{eq:chi2-pder-Vk} yields the best object visibilities given the data and
the complex transmissions:
\begin{equation}
  \label{eq:best-obj-vis}
  V_k^\dagger =
    \frac{
      \displaystyle\sum_{(i,j)\in\mathcal{B}_k}\!
      w_{i,j}\,G_i^\star\,G_j\,\mu_{i,j}
    }{
      \displaystyle\sum_{(i,j)\in\mathcal{B}_k}\!
      w_{i,j}\,\abs{G_i}^2\,\abs{G_j}^2
    }\,.
\end{equation}
Not surprisingly, this solution is a weighted sum of the complex
visibilities measured by sub-aperture pairs which sample the $k$-th
spatial frequency.

The visibilities obtained by Eq.~(\ref{eq:best-obj-vis}) are not normalized.
Assuming $V_0^\dagger$ corresponds to the null frequency, the following
normalization steps insure that the sought visibilities are normalized:
\begin{align}
  \alpha &= V_0^\dagger\\
  V_k^{\dagger} & \leftarrow V_k^{\dagger}/\alpha
  \label{eq:renormalization-of-V}\\
  G_j & \leftarrow \sqrt{\alpha}\,G_j\,.
  \label{eq:renormalization-of-G}
\end{align}
It is worth noting that the likelihood remains the same after these
re-normalization steps.

\subsubsection{Self-calibration stage}
\label{sec:self-calib}

Since, given the complex transmission factors, the best object complex
visibilities can be uniquely derived, the initial optimization problem of
$\ell(\V{V},\V{G})$ can be reduced to a smaller problem which consists in
finding the complex transmissions which minimize the partially optimized
log-likelihood:
\begin{equation}
  \label{eq:reduced-chi2}
  \ell^\dagger(\V{G}) \bydef
  \left.\ell(\V{V},\V{G})\right\vert_{\V{V}=\V{V}^\dagger(\V{G})}
\end{equation}
where $\V{V}^\dagger(\V{G})$ is given by \Eq{eq:best-obj-vis}, possibly after
the re-normalization steps.  The second stage of our algorithm therefore
consists in fitting the complex transmissions so as to minimize
$\ell^\dagger(\V{G})$ with respect to the complex transmissions.

Since the criterion $\ell^\dagger(\V{G})$ is continuously
differentiable, its partial derivatives cancel at any extremum of the
criterion.  Hence the so-called \emph{first order optimality
condition} that at the optimum of $\ell^\dagger(\V{G})$ we must have:
\begin{equation}
  \label{eq:1st-optim-cond-G}
  \frac{\partial\ell^\dagger(\V{G})}{\partial G_i} = 0\,,\quad\forall i\,.
\end{equation}
Note that, unless $\ell^\dagger(\V{G})$ is strictly convex with respect
to the $G_i$'s, \Eq{eq:1st-optim-cond-G} is a necessary condition but is not a
sufficient one because it would be verified by all the extrema (local minima,
local maxima or saddle points) of the criterion.

Since $\V{V}^\dagger$ depends on $\V{G}$, the chain rule must be applied to
derive the partial derivative of $\ell^\dagger(\V{G})$ with respect to the
$i$-th complex transmission.  For instance, the derivative with respect to the
real part of the $i$-th complex transmission expands as:
\begin{eqnarray*}
  \frac{\partial\ell^\dagger(\V{G})}{\partial\Re(G_i)} &=&
  \left.\frac{\partial\ell(\V{V},\V{G})}{\partial\Re(G_i)}
  \right\vert_{\V{V}=\V{V}^\dagger(\V{G})}
  + \sum_k \left.\frac{\partial\ell(\V{V},\V{G})}{\partial\Re(V_k)}
  \right\vert_{\V{V}=\V{V}^\dagger(\V{G})}
  \,
  \frac{\partial\Re(V_k^\dagger)}{\partial\Re(G_i)}\\
  & & \quad
  + \sum_k \left.\frac{\partial\ell(\V{V},\V{G})}{\partial\Im(V_k)}
  \right\vert_{\V{V}=\V{V}^\dagger(\V{G})}
  \,
  \frac{\partial\Im(V_k^\dagger)}{\partial\Re(G_i)}\,.
\end{eqnarray*}
However, since $\V{V}^\dagger$ minimizes $\ell(\V{V},\V{G})$, we have:
\begin{displaymath}
  \left.
  \frac{\partial\ell(\V{V},\V{G})}{\partial V_k}
  \right\vert_{\V{V}=\V{V}^\dagger(\V{G})} = 0\,,
\end{displaymath}
and from the definition in \Eq{eq:cmplx-deriv} of the partial
derivative with respect to a complex variable, we deduce that:
\begin{displaymath}
  \left.
  \frac{\partial\ell(\V{V},\V{G})}{\partial\Re(V_k)}
  \right\vert_{\V{V}=\V{V}^\dagger(\V{G})}
  = 0
  \quad\text{and}\quad
  \left.
  \frac{\partial\ell(\V{V},\V{G})}{\partial\Im(V_k)}
  \right\vert_{\V{V}=\V{V}^\dagger(\V{G})}
  = 0 \,.
\end{displaymath}
It follows that:
\begin{displaymath}
  \frac{\partial\ell^\dagger(\V{G})}{\partial\Re(G_i)} =
  \left.\frac{\partial\ell(\V{V},\V{G})}{\partial\Re(G_i)}
  \right\vert_{\V{V}=\V{V}^\dagger(\V{G})}\,.
\end{displaymath}
Since the same reasoning can be conducted for the derivative with
respect to the imaginary part of the complex transmission and by
definition of the derivation of a real quantity with respect to a
complex variable given in \Eq{eq:cmplx-deriv}, the partial derivative
of the partially optimized log-likelihood finally simplifies to:
\begin{equation}
  %\label{eq:1st-optim-cond-G-alt}
  \frac{\partial\ell^\dagger(\V{G})}{\partial G_i} =
  \left.\frac{\partial\ell(\V{V},\V{G})}{\partial G_i}
  \right\vert_{\V{V}=\V{V}^\dagger(\V{G})}
  \,.
\end{equation}
In words, since $\V{V}^\dagger(\V{G})$ minimizes $\ell(\V{V},\V{G})$
with respect to $\V{V}$, the partial derivative of
$\ell^\dagger(\V{G})=\ell(\V{V}^\dagger,\V{G})$ with respect to
$\V{G}$ is simply the partial derivative of $\ell(\V{V},\V{G})$ with
respect to $\V{G}$ into which the $\V{V}$ is replaced (after
derivation) by $\V{V}^\dagger(\V{G})$.  This property helps to
simplify the calculations to come and, more importantly, shows that
the global optimum must verify the modified first order optimality
condition:
\begin{equation}
  \label{eq:1st-optim-cond-G-alt}
  \frac{\partial\ell^\dagger(\V{G})}{\partial G_i} =
  \left.\frac{\partial\ell(\V{V},\V{G})}{\partial G_i}
  \right\vert_{\V{V}=\V{V}^\dagger(\V{G})}
  =  0\,,\quad\forall i\,.
\end{equation}

Finally, the partial derivative of $\ell^\dagger$ with respect to the
complex transmissions $\V{G}$ can be written:
\begin{eqnarray*}
  \frac{\partial\ell^\dagger(\V{G})}{\partial G_i} &=&
  \left.\frac{\partial\ell(\V{V},\V{G})}{\partial G_i}
  \right\vert_{\V{V}=\V{V}^\dagger(\V{G})}\\
  &=&
  -2\,\sum_k \!\sum_{j:(i,j)\in\mathcal{B}_k}\!\!\!\!
  w_{i,j}\,\left(\mu_{i,j} - G_i\,G_j^\star\,V_k\right)\,G_j\,V_k^\star \\
  && - 2\,\sum_k \!\sum_{j:(j,i)\in\mathcal{B}_k}\!\!\!\!
  w_{j,i}\,\left(\mu_{j,i}^\star - G_i\,G_j^\star\,V_k^\star\right)\,G_j\,V_k \\
  &=& 2\,G_i\,\sum_k \abs{V_k}^2\,\left[
      \sum_{j:(i,j)\in\mathcal{B}_k}\!\!\!\! w_{j,i}\,\abs{G_j}^2
    + \!\!\sum_{j:(j,i)\in\mathcal{B}_k}\!\!\!\! w_{i,j}\,\abs{G_j}^2
      \right] \\
  && -2\,\sum_k \left[
      \sum_{j:(i,j)\in\mathcal{B}_k}\!\!\!\! w_{i,j}\,\mu_{i,j}\,G_j\,V_k^\star
    + \!\!\sum_{j:(j,i)\in\mathcal{B}_k}\!\!\!\! w_{j,i}\,\mu_{j,i}^\star\,G_j\,V_k
      \right]\,.
\end{eqnarray*}
From this last expression, it is tempting to derive a simple iterative
algorithm by solving \Eq{eq:1st-optim-cond-G-alt} for $G_i$ assuming
the other complex transmissions $G_{j:j\not=i}$ are known.  The
resulting recurrence equation is:
\begin{equation}
  G_i^{(n+1)}
  =
  \frac{\displaystyle
    \sum_k \left[
      \sum_{j:(i,j)\in\mathcal{B}_k}\!\!\!\!
      w_{i,j}\,\mu_{i,j}\,G_j^{(n)}\,{V_k^{(n)}}^\star
      + \!\!\!\sum_{j:(j,i)\in\mathcal{B}_k}\!\!\!\!
      w_{j,i}\,\mu_{j,i}^\star\,G_j^{(n)}\,V_k^{(n)}
      \right]
  }{\displaystyle
    \sum_k \Abs{V_k^{(n)}}^2\,\left[
      \sum_{j:(i,j)\in\mathcal{B}_k}\!\!\!\! w_{j,i}\,\Abs{G_j^{(n)}}^2
      + \!\!\!\sum_{j:(j,i)\in\mathcal{B}_k}\!\!\!\! w_{i,j}\,\Abs{G_j^{(n)}}^2
      \right]
  }\,,
  \label{eq:update}
\end{equation}
where $G_j^{(n)}$ is the $j$-th complex transmission at $n$-th
iteration of the algorithm and $V_k^{(n)}$ is the $k$-th best object
visibility computed by \Eq{eq:best-obj-vis} with the complex
transmissions estimated at $n$-th iteration:
\begin{equation}
  \V{V}^{(n)} \bydef \V{V}^\dagger\left(\V{G}^{(n)}\right)
  \,.
\end{equation}

\subsubsection{Multiple exposure case}
\label{sec:selfcal4}

Our algorithm can be generalized to the processing of multiple
exposures of the same object.  We assume that the instrument does not
undergo any significant rotation with respect to the observed object
so that the sampled spatial frequencies (the $\V{u}_k$'s) and the
corresponding sets of sub-aperture pairs (the $\mathcal{B}_k$'s)
remain the same during the total observing time.  We also assume that
the object brightness distribution is stable so that the object
complex visibilities (the $V_k$'s) do not depend on the exposure time.
At least because of the noise and of the turbulence, the measured
complex visibilities and the instantaneous complex amplitude
transmissions however do depend on the exposure index $t$ and are
respectively denoted $\mu_{i,j,t}$ and $G_{i,t}$.  Under the
\citet{1985..Goodman} approximation, the log-likelihood becomes:
\begin{equation}
  \label{eq:chi2-multi}
  \ell(\V{V},\V{G}) = \sum_t \sum_k \sum_{(i,j)\in\mathcal{B}_k}\!\!
  w_{i,j,t}\,\left\vert\mu_{i,j,t} - G_{i,t}\,G_{j,t}^\star\,V_k\right\vert^2
\end{equation}
where the, possibly time dependent, statistical weights are:
\begin{equation}
  \label{eq:weight-multi}
  w_{i,j,t}
  = \frac{1}{\Var\bigl(\Re(\mu_{i,j,t})\bigr)}
  = \frac{1}{\Var\bigl(\Im(\mu_{i,j,t})\bigr)}
  \,.
\end{equation}

Since the object visibilities and the instrumental geometry do not
depend on time, the same spatial frequency is measured at every
exposure by a given pair of sub-apertures.  Hence the condition given
in \Eq{eq:1st-optim-cond-V} can be used to trivially obtain the best
complex visibilities of the object:
\begin{equation}
  \label{eq:best-obj-vis-multi}
  V_k^\dagger =
    \frac{
      \displaystyle \sum_t \sum_{(i,j)\in\mathcal{B}_k}\!
      w_{i,j,t}\,G_{i,t}^\star\,G_{j,t}\,\mu_{i,j,t}
    }{
      \displaystyle \sum_t \sum_{(i,j)\in\mathcal{B}_k}\!
      w_{i,j,t}\,\abs{G_{i,t}}^2\,\abs{G_{j,t}}^2
    }\,,
\end{equation}
which simplifies to \Eq{eq:best-obj-vis} in the case of a single exposure.

The updating formula for the time dependent complex transmissions is
obtained from the condition in \Eq{eq:1st-optim-cond-G-alt} by simply
replacing the aperture index $i$ by an aperture-time index
$i,t$ and straightforwardly:
\begin{equation}
  G_{i,t}^{(n+1)}
  =
  \frac{\displaystyle
    \sum_k \left[
      \sum_{j:(i,j)\in\mathcal{B}_k}\!\!\!\!
      w_{i,j,t}\,\mu_{i,j,t}\,G_{j,t}^{(n)}\,{V_k^{(n)}}^\star
      + \!\!\!\sum_{j:(j,i)\in\mathcal{B}_k}\!\!\!\!
      w_{j,i,t}\,\mu_{j,i,t}^\star\,G_{j,t}^{(n)}\,V_k^{(n)}
      \right]
  }{\displaystyle
    \sum_k \Abs{V_k^{(n)}}^2\,\left[
      \sum_{j:(i,j)\in\mathcal{B}_k}\!\!\!\! w_{j,i,t}\,\Abs{G_{j,t}^{(n)}}^2
      + \!\!\!\sum_{j:(j,i)\in\mathcal{B}_k}\!\!\!\! w_{i,j,t}\,\Abs{G_{j,t}^{(n)}}^2
      \right]
  }\,,
  \label{eq:update-multi}
\end{equation}
which also simplifies to \Eq{eq:update} in the case of a single exposure.

\subsubsection{Algorithm summary}
\label{sec:selfcal5}

Putting everything together, our algorithm consists in the following steps:
\begin{enumerate}
\item initialization: set $n=0$ and choose the starting complex
  transmissions $\V{G}^{(0)}$;
\item compute the best object visibilities $\V{V}^{(n)}$ given the complex
  transmissions $\V{G}^{(n)}$ according to \Eq{eq:best-obj-vis-multi} and,
  optionally, renormalize the unknowns;
\item terminate if the algorithm converged; otherwise, proceed with next step;
\item compute $\V{G}^{(n+1)}$ by updating the complex transmissions
  according to \Eq{eq:update-multi};
\item let $n:=n+1$ and loop to step 2;
\end{enumerate}

Our iterative algorithm is very simple to implement and its modest
memory requirements makes it possible to process over several
thousands of snapshots all together. This is a requirement for faint
objects or to achieve very high dynamic range.  Yet, on a strict
mathematical point of view, our algorithm may have a number of
deficiencies.  First, as already mentioned, the first order optimality
condition is necessary but not sufficient to insure that the global
minimum (or even a local minimum) of $\ell^\dagger(\V{G})$ has been
reached.  Other non-linear image reconstruction algorithms (blind
deconvolution, optical interferometry imaging, ...) have the same
restriction. In practice, checking that the algorithm converges toward
a similar solution for different initial conditions can be used to
assert the effective robustness of the method.  A second possible
problem results from the updating of the complex transmission by
\Eq{eq:update-multi}.  If a fixed point is discovered by the
recursion, then it satisfies the necessary optimality condition; but
it is also possible that the recursion gives rise to oscillations for
the values of the sought parameters.  Note that our updating method is
a non-linear one, analogous to numerical methods for solving linear
equations such as the Jacobi and Gauss-Seidel methods
\citep{templates}.  Unlike for our specific problem, it is however
possible to prove the convergence of the recursion in the linear case.
Again, the behavior of the algorithm in practice can effectively prove
its ability to converge to a fixed point.  If it appears that the
update formula leads to oscillations, this problem can be completely
solved by using an iterative optimization algorithm which guarantees
that the partially optimized log-likelihood $\ell^\dagger(\V{G})$ is
effectively reduced from one iteration to another.  Since the
log-likelihood is a sum of squares, a Levenberg-Marquardt algorithm
\citep{More1977LevenbergMarquardtAlgorithm} coupled with a trust
region method \citep{More_Sorensen-1983-trust_region_step} would
completely solve this problem.  In practice, none of the numerous
simulations we have conducted with our iterative algorithm have given
rise to any of these convergence problems.

Although derived in the specific case of a pupil remapping instrument,
our algorithm shares similarities with the self-calibration method
used in radio-astronomy \citep{1981MNRAS.196.1067C}.  However, in our
case, not only the phases of the complex transmissions are
miscalibrated and must be recovered but also the amplitudes.  Besides,
we do not need a regularization term to overcome the sparsity of the
$(u,v)$ coverage by radio interferometers.  Our derivation of the
non-linear updating formula in \Eq{eq:update-multi} is also quite similar
to the iterative method proposed by \citet{Matson1991:phase} for
recovering the Fourier phase from the bispectrum phase and later
improved by \citet{Thiebaut1994:PhD} to achieve better convergence
capabilities.

\section{Dynamic range estimations} \label{sec:dyna}

\subsection{Analytical estimation of the photon noise limitations}
\label{sec:analy}

This system gives calibrated measurements of the spatial frequencies
of the object. The advantage is straightforward. In classical imaging,
phase and amplitude errors create speckles in the image plane,
therefore limiting the dynamic range. With a pupil remapped
instrument, and assuming we are acquiring fast enough to {\it freeze}
atmospheric turbulence, statistical errors due to photon and detector
noise will theoretically be the main limiting
factor. \citet{2002.Baldwin} showed that the dynamic range of a
reconstructed image is linked to the errors of the Fourier components:
\begin{equation}
  \dyn = \sqrt{\frac{n}{(\delta V/V)^2+(\delta \phi)^2}} \, ,
\end{equation}
where $n$ is the total number of data points, $(\delta V/V)$ is the
fractional error in amplitude, and $\delta \phi$ the phase error (in
radians). For a total number of photons $\Nph$ and a number of
apertures $M$, the amplitude of the fringe peaks in the Fourier
transform of the image is equal to $\Nph/M$ (assuming full coherence
for the fringes). Considering a white photon noise of amplitude
$\sqrt{\Nph}$, the signal-to-noise of the visibility modulus can be
estimated with
\begin{equation}
  V / \delta V = \frac{\sqrt{\Nph}}{M} \, ,
\label{eq:sb_pour_V}
\end{equation}
as for the phase \citep{1985..Goodman}:
\begin{equation}
\delta \phi \approx \frac{\delta V}{ V} \, .
\end{equation}
This leads to the following approximation of
the dynamic range:
\begin {equation}
\dyn = \sqrt{\frac{M(M-1)}{2 M^2/\Nph}} \approx \sqrt{\frac{\Nph}{2}} \, .
\label{eq:dyn}
\end{equation}
Within these approximations, this result has the merit of clearly
highlighting the advantage and the drawback of a single-mode
remapping system: (i) an arbitrarily high dynamic range can be
obtained anywhere in the image, providing the integration time is long
enough; (ii) since additive noise is uniformly distributed on all the
spatial frequencies, it is also evenly distributed across the whole
field of view. To compare, an optical design isolating the photons of
a bright object next to a faint companion -- like a perfect adaptive
optics and coronographic system -- would achieve a superior
photon-wise dynamic range of $\dyn=\Nph$. Extreme dynamic range
imaging, as required for detecting extra-solar earths
($\dyn\approx10^{10}$), would therefore also require a long integration
time with our system.

\subsection{Numerical simulations}

\begin{figure}
  \centering{\includegraphics[width=5.5cm]{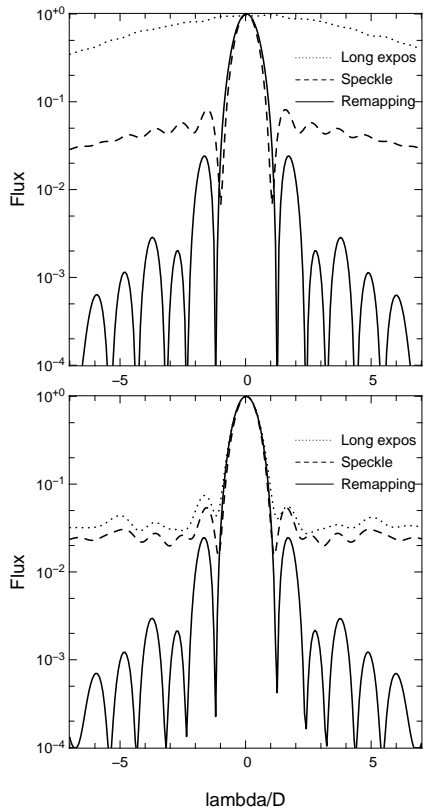}}
  \caption{ \label{fig:Ao_No_Comp2} Horizontal cuts of the point
    spread functions imaged in \Fig{fig:Ao_No_Comp}. The source is a
    star of magnitude 5, observed at 630 nm.  The acquisition setting
    consists of 10\,000 snapshots of 4 ms each on an 8 meter telescope
    ($r_0 \approx 20$ cm). The upper panel shows the point spread
    function after an uncorrected turbulence, while the lower panel
    shows the PSF after partial correction of the wavefront by an
    adaptive optics system. The three curves are obtained through the
    Eqs.~(\ref{eq:lp}), (\ref{eq:sp}) and (\ref{eq:rp}). The last
    equation require a pupil remapping to get an estimation of the
    $\widetilde{G}_i$ and $\widetilde{G}_j$ terms. At a few resolution
    elements from the star, AO systems and speckle techniques cannot
    achieve a dynamic range over 40. However, pupil remapping enables a
    perfect reconstruction of the PSF, with dynamic ranges over
    $10^3$.  }
\end{figure} 

\subsubsection{Simulation setup}

To perform these simulations, we used ``YAO'', an adaptive optics
simulation software written by F.~Rigaut using the Yorick
language. This software allows us to generate corrugated wavefront with
and without adaptive optics correction.  In our simulations, the
instrumental setup corresponds to an 8 meter telescope under good
seeing condition ($r_0\approx20\ \mathrm{cm}$ at $630\ \mathrm{nm}$)
caused by four different layers of turbulence at altitudes 0, 400,
6\,000 and 9\,000 meters. The wind speed ranges from 6 to $20\
\mathrm{m/s}$ depending on the altitude of the layer. The AO system is
optimized to work in the near infrared. It consists in a classical
Shack-Hartmann wavefront sensor and a $12\times12$ actuator deformable
mirror. The loop frequency has been set to $500\ \mathrm{Hz}$, with a
gain of 0.6 and a frame delay of 4 ms. The guide star is of magnitude
5.

The remapping was done by dividing the 8\ meter telescope pupil into
132 hexagonal sub-pupils of 80\ centimeters in diameter each. They are
filtered by the fundamental mode of single-mode fibers, coupled
so as to maximize the injection throughput of an uncorrupted
incoming wavefront. The injection efficiency in this case would be of
78\%. However, at the operating wavelength of $630\ \mathrm{nm}$, the
diameters of the sub-pupils are large compare to the Fried parameter
($d/r_0 \approx 4$) and the coupling is expected to be much lower
without adaptive optics ($\approx 5\%$ in these simulations). The 132
sub-pupils are then rearranged in a non-redundant configuration, to
produce a total of 8\,646 sets of fringes on the detector.

The total integration time was set to 40\ seconds. However, because of
the coherence time of the atmosphere, acquisition was done by
sequences of short acquisition periods. We used a snapshot time of 4
milliseconds, during which we integrated the effects of phase
variations. This was a way to account for fringe blurring due to
dynamic piston effects. We also added to our measurements the photon
noise as a Gaussian noise of variance the number of photons on each
pixel. The number of photons was computed to account for a
coupling efficiency of 5\% into the fibers and a spectral
bandpass of $60\ \mathrm{nm}$.

No chromatic fringe blurring was introduced since its influence
would highly depend on the chosen technical setup. Moreover,
there are several ways to avoid this problem. In the case of a 1D
non-redundant remapping, we recommend to spectrally disperse the
fringes. If a 2D non-redundant reconfiguration is mandatory due to a
large number of sub-pupils, another solution could be to use a
hyper-chromatic magnifier like a Wynne lens system \citep[as proposed
by][]{2004SPIE.5491.1624R}.  At least, the use of a narrow spectral
filter can completely avoid the chromatic blurring.

%Finally a last possibility is to use a narrow spectral bandpass.

\subsubsection{Comparison with the speckle and adaptive optics techniques}
\label{sec:Comp_Ao_No}

\begin{figure*}
  \centering{\includegraphics[width=13cm]{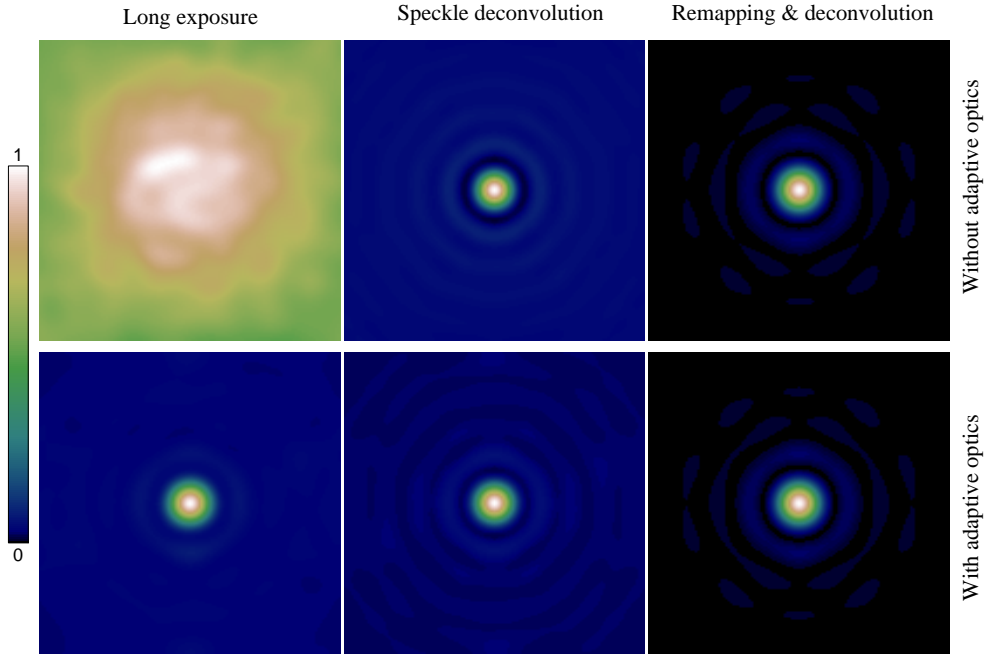}}
  \caption{ \label{fig:Ao_No_Comp} Point spread functions (PSF) of the
    instrument using different imaging techniques. The
    point-like object is a star of magnitude 5, observed at 630
    nm.  The acquisition setting consists of 10\,000 snapshots of 4 ms
    each on a 8 meter telescope. {\em Left panels:} average of the
    exposures over the total acquisition time (cf.\ \Eq{eq:lp}). {\em
    Central panels:} reconstruction obtained by using
    conventional speckle interferometry technique (cf.\ \Eq{eq:sp}).
    {\em Right panels:} PSF after remapping and correction by its
    estimation (cf.\ \Eq{eq:rp}). {\em Upper panels:} the PSF is obtained
    trough an atmospheric turbulence of $r_0 \approx 20$ cm ($D/r_0
    \approx 40$). {\em Lower panels:} PSF with the same corrugated
    wavefront but corrected by a simulated adaptive optics system.
    The field of view of each image is around 15 resolution elements
    ($15 \lambda/D$). The PSF quality goes from bad (long exposure
    without adaptive optics), to medium (AO correction and/or Speckle
    deconvolution), to perfect (after a remapping and a post-detection
    processing).}
\end{figure*}

This first test was performed in order to demonstrate the
reconstruction quality of the PSF.  The astronomical object is a
point-like source of Fourier transform values 1 ($V_k = 1,\ \forall
k$). The observing wavelength is 630 nm.  As described in the previous
section, the simulated dataset consists of 10\,000 snapshots, each
featuring 8646 fringe sets. From each set of fringes, a complex
coherence value $\mu_{i,j}$ is extracted, and complex transmission
factors $\widetilde{G}_i$ are estimated according to the
iterative algorithm described in Sect.~\ref{sec:self-calib}. Finally,
we used \Eq{eq:rp} to obtain the calibrated OTF, and thus the
PSF. For comparison, the same corrugated wavefronts were used to
obtain the PSF with two other techniques.  The first one consists in
averaging the complex instantaneous OTF by \Eq{eq:lp}.  The second one
(speckle interferometry) consists in averaging the squared modulus,
removing photon noise bias and taking the square root as in
\Eq{eq:sp}.  In this work, we did not make use of the bispectrum or
closure phase since our object is point-like and therefore purely
symmetric. The four left panels of \Fig{fig:Ao_No_Comp} represent the
deduced PSF, with (lower panels) and without (upper panels) the use of
an adaptive optics system.

The first result confirm the usefulness of both speckle interferometry
and adaptive optics systems, even though we are observing at visible
wavelengths. This is clearly seen in \Fig{fig:Ao_No_Comp2}; where
unlike uncorrected long exposure imaging, they permit the retrieval of
spatial information at the diffraction limit of the
telescope. Nevertheless, the Strehl ratio in images obtained by these
techniques is very low and the background pollution remains
important. Without remapping, the best dynamic range achievable at
visible wavelength is with a combination of AO and speckle
interferometry technique, which provide a dynamic range of 40. This of
course is in the case of a perfectly working near-infrared AO system
with a 500 Hz frequency loop. Most current AO system are not used in
that mode since any slight miss-calibration of the actuator influence
function would make it useless at these wavelengths.

On the other hand, a diffraction pattern is obtained by using our
remapping instrument and our image reconstruction algorithm. The
pattern differs slightly from a perfect Airy disc due to the hexagonal
sampling of the OTF.  At a few $\lambda/D$ from the central
star, a dynamic range of $10^3$ is obtained (see
\Fig{fig:Ao_No_Comp2}).  Using an adaptive optics system does
not significantly modify the pattern, meaning that the
dynamic range is limited by the shape of the perturbation free PSF.
Our conclusion is therefore that the dynamic range could be further
increased by choosing a different PSF. We did this in the following
section over a complex astronomical object.

\subsubsection{Image reconstruction}
\label{sec:complex}

In this simulation, the astronomical object is a star surrounded by a
protoplanetary disc. The disc has an exponential brightness
distribution and a total flux of a hundredth of the star
flux. Two companions are also present, one with a flux of a
thousandth, and the other of ten thousandths the flux of the star.

As before, the $G_i$ complex transmissions are estimated by using
iteratively \Eq{eq:update}. But unlike in Sect.~\ref{sec:Comp_Ao_No},
calibrated object visibilities are retrieved by our algorithm
using \Eq{eq:best-obj-vis-multi} from 10\,000 short exposure
images. We used these visibilities to reconstruct an image. But
instead of doing a simple Fourier transform to obtain the equivalent
of a {\em dirty map}, it is possible to choose the PSF according to a
scientific goal. For example, an Airy disc PSF is good to achieve high
angular resolution, while a Gaussian apodization is better for the
dynamic range. Of course, any other visibility apodization can also be
applied to obtain the most suitable PSF. This system thus has the
advantage of producing images close to what can be obtained with
optically apodized pupil optics
\citep{2002ApJ...570..900K,2003A&A...404..379G}, while completely free
of any atmospheric perturbations. This way, the dynamic range is no
longer limited by the Airy rings of the diffraction pattern of the
telescope.

\begin{figure*}
  \centering
  \resizebox{\hsize}{!}{\{\includegraphics{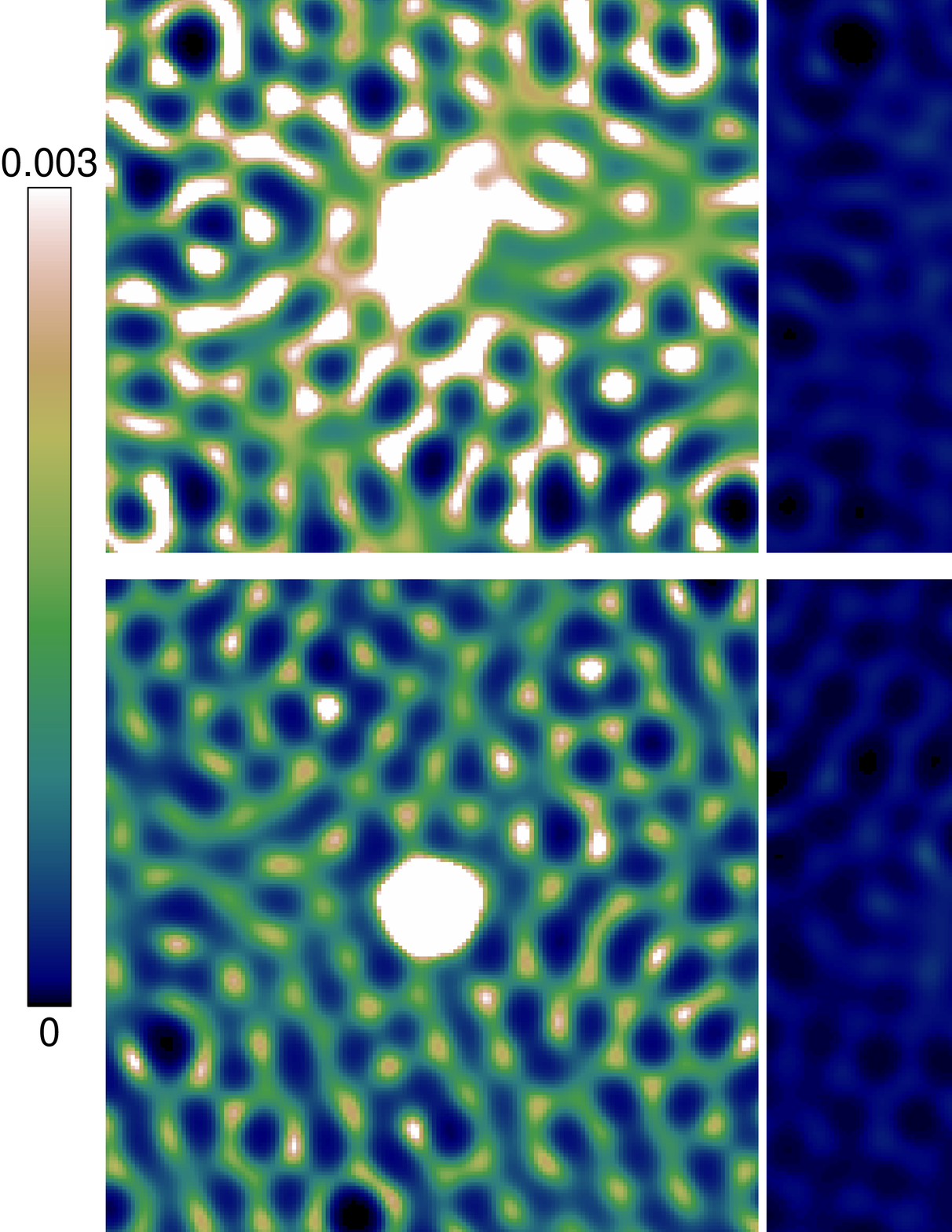}}
  \caption{ \label{fig:figure_complex} These simulations feature
    reconstructed image from the visibilities acquired from
    \Eq{eq:best-obj-vis}. The acquisition settings consist of
    10\,000 snapshots of 4 ms each on an 8 meter telescope in the
    visible ($r_0 \approx 20$ cm). The object is a central star with a
    protoplanetary disc and two companions, of relative flux a
    thousandth and ten thousandths (respectively, at the upper-right
    and upper-left of the central star -- they are highlighted by the
    red circles on the upper right image). {\em From left to right},
    the difference in the reconstructed images are due to the
    brightness of the object (magnitudes 15, 10, 5 and 0). {\em The
    upper panels} are reconstructed images in the case of an
    uncorrected wavefront ($D/r_0 \approx 40$). {\em The lower panels}
    are reconstructed images with an AO corrected wavefront. The field
    of view of each image is around 30 resolution elements ($30
    \lambda/D$). The color scale on the right is linear, from 0 to $3
    \times 10^{-3}$, and normalized to the flux of the central star.
    The reconstructions show dynamic range around or over $10^4$,
    except for the ones with a star of magnitude 15. In the three
    upper panels, we can clearly see the photon noise limit, evolving
    as a function of the brightness of the source.  For faint sources,
    the lower-right panel show that dynamic range is increased by
    combining a remapping setup with an adaptive optics
    system. However, on a bright source, fringe smearing due to the
    high frequency control loop of the deformable mirror limit the
    dynamic range. An horizontal cut of these figures can be seen in
    \Fig{fig:figure_complex2}. }
\end{figure*}

\begin{figure}
  \centering
  {\includegraphics[width=5.5cm]{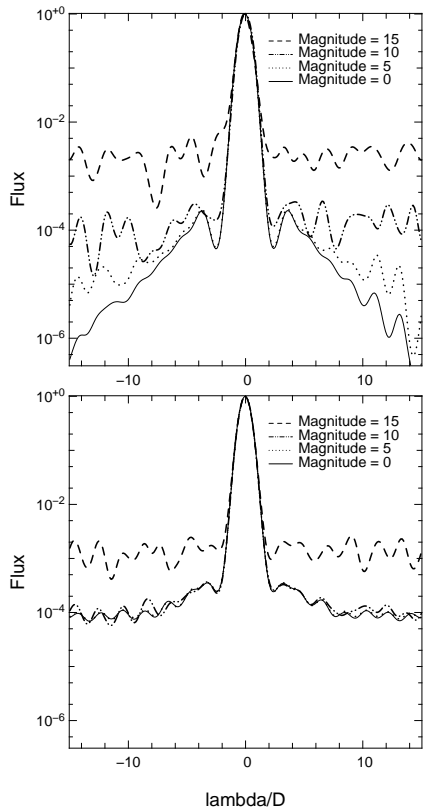}}
  \caption{ \label{fig:figure_complex2} Horizontal cuts of the
    reconstructed images showed in \Fig{fig:figure_complex}.  The
    acquisition setting consists of 10\,000 snapshots of 4 ms each on
    an 8 meter telescope ($r_0 \approx 20$ cm). The upper panel shows
    the reconstruction in the case of an uncorrected turbulence, while
    the lower panel shows the reconstruction with adaptive optics
    correction. On the upper panel, we can clearly see the dynamic
    range improving with the brightness of the source, from $10^3$ (15
    mag), to $10^6$ (0 mag). At the highest dynamic range, we can see
    the exponential brightness decrease of the protoplanetary disc.
    Behind an adaptive optic, deconvolution is no longer limited by the
    photon noise, but by the high frequency differential piston
    introduced by the deformable mirror.  }
\end{figure}

However, there is one perturbation this system cannot correct for; it
is the photon noise. The phase corrugation being accounted for, the
photon noise should be the theoretical limit of the dynamic range.  To
try to reach this limit, we used a Gaussian shape apodization on the
visibilities. All of the image reconstructions in
\Fig{fig:figure_complex} are of the object described in the first
paragraph of this section, but with different total brightnesses. The
reconstructions from the left to the right are respectively of a
central star of magnitude 15, 10, 5, and 0. The upper set of panels
are for a system without AO, while the four bottom panels are with AO
activated. \Fig{fig:figure_complex2} gives a summary by plotting
a horizontal cut of the different reconstructions.  Table~\ref{tb:dyn}
lists the dynamic ranges estimated on the images and compares them with
the analytical approximation of the photon noise established in
Sect.~\ref{sec:analy}. The dynamic range estimations are obtained by
taking the inverse of the root mean square of the background of the image
normalized by its pixel of maximum flux.

\begin{table}
\caption{Dynamic range results}
\label{tb:dyn}
\centering
\begin{tabular}{c c c c c}
\hline \hline
 &\multicolumn{2}{c}{Without AO} &\multicolumn{2}{c}{With
  AO} \\
Magnitude & $\sqrt{\Nph/2}^{\mathrm{a}}$ & D.R.$^{\mathrm{b}}$ &
 $\sqrt{\Nph/2}^{\mathrm{a}}$ & D.R.$^{\mathrm{b}}$ \\
\hline
0 & $1.1 \times 10^6$ & $0.9 \times 10^6$ & $2.4 \times 10^6$ & $1.8 \times 10^4$   \\
5 & $1.1 \times 10^5$ &  $1.5 \times 10^5$ & $2.4 \times 10^5$ & $1.7 \times 10^4$   \\
10 & $1.1 \times 10^4$ &  $1.3 \times 10^4$ & $2.4 \times 10^4$ & $1.6 \times 10^4$  \\
15 & $1.1 \times 10^3$ & $0.8 \times 10^3$  & $2.4 \times 10^3$ & $1.2 \times 10^3$   \\
\hline
\end{tabular}
\begin{list}{}{}
\item[$^{\mathrm{a}}$] Theoretical dynamic range as predicted by \Eq{eq:dyn}.
\item[$^{\mathrm{b}}$] Dynamic range obtained by taking the
  background rms of the reconstructed images of \Fig{fig:figure_complex}.
\end{list}
\end{table}

When atmospheric turbulence is not corrected by an AO system, the
reconstructions clearly highlights a dynamic range limited by photon
noise. For an object of magnitude 10, we achieved a dynamic range of
$1.3 \times 10^4$. This is comparable to what was calculated from the
total photon count and the analytical relation $\dyn \approx
\sqrt{\Nph/2}$. This theoretical value, taking into account the 5\%
coupling efficiency in the fibers, was of $1.1 \times 10^4$. A second
interesting point is that when the brightness of the source increases
by a factor 100 (delta mag = 5), the dynamic range increases by a
factor around 10. This can be seen over 15 order of magnitude,
indicating that the dynamic range has a roughly linear increase as a
function of $\sqrt{\Nph}$. This does 1) confirm the validity of the
analytical estimation, and 2) prove the quality of our self
calibration algorithm to restore the object visibilities.

Noteworthy is the reconstruction of the system with a brightness of
15 mag. Due to a relatively low coupling efficiency in the fibers, an
average of 250 photons are detected per 4 ms snapshot. According to
\Eq{eq:sb_pour_V}, it implies a S/N ratio of $\approx 0.12$ for each
spatial frequency measurement. It is therefore impressive to note that
even with such a vague knowledge of the $\mu_{i,j}$, the iterative
algorithm of Sect.~\ref{sec:self-calib} is capable of restoring images
with a dynamic range limited by the photon noise. This is possible
since it combines the advantage of fitting the information from all
snapshots together with a small, cpu friendly, recursive algorithm.

When atmospheric turbulence is corrected by an AO system, the
dynamic range does not increase linearly as a function of the
brightness anymore. Indeed, in the lower panels of
\Fig{fig:figure_complex} and \Fig{fig:figure_complex2}, the dynamic
range is clearly limited by another factor. Thanks to the correction,
the injection throughput is higher, around 23\%. This allows an
increase in dynamic range for the faintest source, but with a
saturation around a few $10^{4}$. The reason for this unpredicted
threshold appeared clearly when analyzing our simulations. During the
4 ms integration time, mirror displacements happened twice to adapt for
the atmospheric turbulence. These minor corrections, while increasing
the spatial coherence \citep{2000OSAJ...17..903C}, have the
drawback of introducing a phase noise overshoot
at the frequency of the loop.  This side effect results in a slight
blurring of the fringes recorded on the detector and a bias on the
measurements. To profit from the advantages of the AO, solutions could
either be to increase the acquisition rate, or, better, to adjust the
control loop to decrease the amplitude of the overshoot.

\section{Summary}
\label{sec:conc}

In this paper, we presented further investigation of the
instrument introduced in \citet{2006..Perrin}.

\begin{itemize}

\item We established an analytical relation linking the Fourier
components of the image, the Fourier components of the object, and the
atmospheric perturbations (\Eq{eq:mul_l}). We showed that
inverting this equation allowed us to completely disentangle turbulence
from astronomical information (Sect.~\ref{sec:why_remap}).

\item We developed an analytical iterative self-calibration algorithm
which enables inversion of the previously established equation over
several thousands of snapshots simultaneously. This algorithm happened
to be robust, allowing visibility determination from Fourier component
measurements with S/N ratio well below 1 (Sect.~\ref{sec:selfcal}).

\item Simulations of this system confirmed the validity of the
algorithm and produced high dynamic range, diffraction limited, images
of complex astronomical objects. A dynamic range of the order of
$10^6$ was achieved at visible wavelength on an eight meter telescope
and in the presence of good seeing conditions. Compared to actual AO
systems, it represents an increase of around $10^4$ in dynamic range.
We noted that the sensitivity of the instrument would increase by
using an adaptive optics system, but at the price of a limitation in
the achievable dynamic range (Sect.~\ref{sec:dyna}).

\end{itemize}

%%%%%%%%%%%%%%%%%%%%%%%%%%%%%%%%%%%%%%%%%%%%%%%%%%%%%%%%%%%%%

\section*{Acknowledgments}
  The authors would like to thank F.~Rigaut for letting ``YAO'' freely
  available to the astronomical community. ``YAO'' is a simulation
  tool for adaptive optics systems, available on the web site
  http://www.maumae.net/yao/. Simulations and data processing for this
  work have been done using Yorick language which is freely available
  at http://yorick.sourceforge.net/. The authors would also
  like to thank the many people with whom discussing helped a lot to
  mature this paper. These persons include F.~Assemat, P.~Baudoz,
  A.~Bocalletti, V.~Coud\'e du Foresto, E.~Ribak, G.~Rousset and
  P.~Tuthill.

%%%%%%%%%%%%%%%%%%%%%%%%%%%%%%%%%%%%%%%%%%%%%%%%%%%%%%%%%%%%%
%%%%% References %%%%%

\bibliographystyle{mn2e}   %>>>> makes bibtex use aa.bst
\bibliography{pupil_remapping_imaging}   %>>>> bibliography data in report.bib

\end{document}